\documentstyle[prl,aps]{revtex}
\begin{document}
\advance\textheight by 0.5in
\advance\topmargin by -0.25in
\draft

\twocolumn[\hsize\textwidth\columnwidth\hsize\csname@twocolumnfalse%
\endcsname

\preprint{NSF--ITP--95--132, cond-mat/9510089}

\title{ {\hfill\normalsize NSF--ITP--95--132, cond-mat/9510089\medskip\\}
Chiral Surface States in the Bulk Quantum Hall Effect}

\author{Leon Balents and Matthew P. A. Fisher}
\address{Institute for Theoretical Physics, University of California,
Santa Barbara, CA 93106--4030}

\date{\today}

\maketitle

\begin{abstract}
In layered samples which exhibit a bulk quantum Hall effect
(QHE), a two-dimensional (2d) surface ``sheath" of gapless
excitations is expected.  These excitations comprise
a novel 2d chiral quantum liquid which should dominate
the low temperature transport along the field (z-axis).
For the integer QHE, we show that localization effects
are completely absent in the ``sheath", giving a metallic
z-axis conductivity.
For fractional filling $\nu =1/3$, the ``sheath"
is a 2d non-Fermi liquid, with incoherent z-axis transport and
$\sigma_{zz} \sim T^3$.
Experimental implications for the Bechgaard salts are discussed.

\end{abstract}
\vskip -0.5 truein
]

Disorder has a profound effect on transport in two-dimensional (2d)
electron systems.\cite{Lee}.  In the absence of an applied magnetic
field, all the electronic states are believed to be localized due to
strong quantum interference effects.  With weak disorder, when the
conductivity $\sigma$ is larger than $e^2/h$, there is a
weak-localization regime described by diffusive behavior with
logarithmic temperature corrections.  In the low temperature limit,
however, a crossover to strongly localized behavior with $\sigma \ll
e^2/h$ is always expected.  In this regime the conductivity drops
rapidly with temperature and vanishes as $T \rightarrow 0$.

In the presence of a magnetic field, 2d localization can be
circumvented by tuning to the center of Landau levels.  At these
isolated transitions between quantum Hall plateaus, extended states
and a temperature independent conductivity of order $e^2/h$ are
predicted, consistent with experiment\cite{Bodo}.  There are also
other 2d systems which exhibit such ``metallic" behavior at isolated
transitions, for example 2d films at the superconductor-insulator
transition.  But away from such transitions, when the conductivity is
well below $e^2/h$, it is invariably strongly temperature dependent,
and insulating at zero temperature.

In this paper, we describe a novel class of anisotropic 2d electronic
phases, which surprisingly can exhibit ``metallic" conductivities much
smaller than $e^2/h$.  These 2d phases arise at the {\it surface} of
bulk {\it three}-dimensional quantum Hall samples.  We consider
layered samples which exhibit independent quantum Hall states in each
layer when a large perpendicular magnetic field is applied.  This
requires an inter-layer tunneling amplitude $t$ small compared to the
2d quantum Hall gap $E_g$.  In such materials, the surface of the
sample is enveloped by a sheath of current-carrying states as depicted
in Figure 1.  This chiral
surface phase is the 2d analog of the 1d states at the edges of a
single layer quantum Hall fluid.  There are currently two candidate
experimental realizations of such systems: 2d electron gas
multi-layers, and the Bechgaard salts, (TMTSF)$_2$X (where TMTSF is
tetramethyltetraselenafulvalene, and X = PF$_6$, ClO$_4$, or ReO$_4$),
which exhibit simultaneous field-induced spin density waves (SDWs) and
QHE\cite{SDWexpts,SDWtheory}.

Electronic transport parallel to the field
direction is a powerful probe of
these chiral surface phases.
Our analysis reveals that for the integer quantum Hall effect (IQHE),
the motion along the z-direction is always diffusive,
independent of the impurity scattering strength.
Localization effects - weak or otherwise - are
completely absent.
The 2d resistivity $\rho_{zz}$ is temperature
independent as $T \rightarrow 0$, with a value which can
be much {\it larger} than $h/e^2$.  But perpendicular
to the field, along the x-axis, the transport is ballistic
and the resistivity $\rho_{xx}$ vanishes as $T \rightarrow 0$.
Our predictions for various experimental quantities
which should be accessible in
the Bechgaard salts, are given in
detail at the end of the paper.

The surface sheath in the fractional quantum Hall effect
at filling $\nu =1/3$ is even {\it more} anisotropic.
In this case, we find that electron transport along the z-axis
is always incoherent.  At low temperatures
insulating behavior is predicted with a
resistivity diverging as $\rho_{zz} \sim 1/T^3$,
even in the absence of any impurity scattering.
At $T=0$ inter-layer transport
is completely absent, and the electrons are ``confined" to 1d.
This chiral surface sheath for $\nu=1/3$ is a nice example
of a 2d non-Fermi liquid phase.

\begin{figure}[hbt]
\vspace{15pt}
\caption{(a): Geometry of a 3d quantum Hall sample.  $z$--axis
transport is included via the tunneling amplitude $t$, and impurity
scattering by the random potential $V$.  (b): Associated Fermi surface
for the sheath.  It differs from that of a conventional open
Fermi-surface metal by the absence of the left-moving (dotted) half.}
\label{fig1}
\end{figure}

Consider first the IQHE.  The electronic states at the edge can be
modeled simply as chiral Fermions\cite{Halperin_edge}.  In the 3d case of
interest here, there is a single edge state per layer ($x$--$y$ plane) per
filled Landau level (see Fig.~1).  We focus on the case $\nu=1$; higher
integer filling fractions behave similarly.  Including a small
inter-layer matrix element $t$ leads to the non-interacting
Hamiltonian
\begin{eqnarray}
H & = & \sum_i \int\! dx \psi_i^\dagger i v \partial_x \psi_i -
t\left(\psi_i^\dagger \psi_{i+1} + \psi_{i+1}^\dagger\psi_i\right)
\label{H0x}\\
& = & \int_{\bbox{p}}\! \left(vp_x - 2 t \cos
p_z\right) \psi^\dagger(p_x,p_z) \psi(p_x,p_z),
\label{H0p}
\end{eqnarray}
where $\int_{\bbox{p}} \equiv \int_{-\infty}^\infty
dp_x/(2\pi)\int_{-\pi}^\pi dp_z/(2\pi)$ and $\psi^\dagger_i$ is an
electron creation operator for the edge state in the $i^{th}$ layer.
Here we have set the layer spacing $a$ equal to one.
Eq.~\ref{H0p}\ shows that inter-layer
hopping induces a small dispersion along the layering direction,
resulting in ``half'' of an open Fermi surface (see Fig.~1b).

Before including impurity scattering, we re-examine
the validity of Fermi-liquid theory (FL)
with electron-electron interactions present.
Despite the chiral nature of the sheath,
the phase-space restrictions which stabilize the FL
for a conventional Fermi-surface
continue to apply here.  In
particular, the quasi-particle decay rate ${\rm Im}\Sigma(\omega) \sim
\omega^2 \ln\omega$, the d.c. conductivity $\sigma_{zz} \sim 1/T^2$,
and the specific heat $C \sim  T/(av)$.  Long-range Coulomb interactions
are screened in the static ($\omega \rightarrow 0$) limit.

To study d.c. transport at low temperatures, we must include impurity
scattering at the surface, which we include by adding to the Hamiltonian
a term of the form,
\begin{equation}
H_{imp} = \sum_i \int\! dx V(x,i a) \psi_i^\dagger
\psi_i .
\end{equation}
For simplicity, we assume the random potential
to be Gaussian and uncorrelated, $\overline{V(x,z)V_j(x',z')} = \Delta
\delta(x-x')\delta_{z,z'}$.
In contrast to the results for the pure
system, the chiral nature of the dirty sheath leads to dramatic
differences from ordinary dirty 2d metallic behavior.
In particular,
localization along the x-direction is precluded {\it a priori}
since the wavefunctions $\Phi$ are necessarily de-localized
along the $x$-direction.  This follows from the
continuity equation $\partial_x |\Phi|^2 =
\nabla_z \cdot J$, where $J(x,z) = (2t/v) {\rm
Im}[\Phi^*(x,z)\Phi(x,z+a)]$, valid for arbitrary random potential $V$.
To study possible localization effects along the z-axis,
we consider the usual averaged product of advanced
and retarded Green's functions for non-interacting electrons,
\begin{equation}
{\cal D}({\bf r};\Omega) \equiv
\overline{G_+(\bbox{0},\bbox{r};E)G_-(\bbox{r},\bbox{0};E+\Omega)} .
\label{Kdef}
\end{equation}
Here $G_\pm = [ iv\partial_x - t\nabla_z^2 -E \mp i \eta +
V(\bbox{r})]^{-1}$, with $\nabla_z^2$ the discrete Laplacian, and
$\eta$ a positive infinitesimal.  A standard summation of ladder
diagrams, which gives a diffusive form for non-chiral Fermions, yields
the approximation
\begin{equation}
{\cal D}_{\rm ladder}(p_x,p_z;\Omega) = {{2\pi\rho} \over {-i(\Omega + vp_x)
+ D p_z^2 + 2\eta}},
\label{laddersum}
\end{equation}
where the density of states $\rho = 1/{2\pi v a}$, and $D= t^2\tau
a^2$, with the relaxation time $\tau = 2v/\Delta$.  The anisotropic form of the
denominator in
Eq.~\ref{laddersum}\ reflects the difference in propagation along $x$
(ballistic) and $z$ (diffusive).  The Einstein relation $\sigma_{zz} =
e^2 \rho D$ can be verified by explicit computation.
The derivation of the diffusive transverse motion in Eq.~\ref{laddersum}\ has
neglected possibly
important quantum interference corrections.  Indeed,
for isotropic 2d systems, such corrections invalidate
diffusive behavior at low temperatures, due to localization.
To study possible interference effects here, we use the $Q$--matrix
approach, which allows a systematic treatment of corrections to the
simple ladder result in Eq.~\ref{laddersum}.  Following McKane and
Stone\cite{MS}, averaged correlation functions are obtained from the
replicated partition function $Z = \int [d\bar{\phi}][d\phi][dQ]
\exp(-S)$, where
\begin{eqnarray}
S & = & \sum_i \int \! dx \bigg\{ i\bar{\phi}\Lambda\left[iv\partial_x
- t\nabla_\perp^2
- E - \Delta Q\right]\phi \nonumber \\
& & + {\Delta \over 2} {\rm Tr} Q^2 + \eta \bar{\phi}\phi \bigg\}.
\label{pf1}
\end{eqnarray}
Here $\phi$ is a $2n$-component complex vector, $Q$ is a $2n$ by $2n$
hermitian matrix, and $\Lambda = 1 \otimes \sigma_z$.  Eq.~\ref{pf1}\
follows from introducing $\phi_\pm$ fields to generate retarded and
advanced Green's functions, averaging the $n^{\rm th}$ moment of the
original generating function $Z_0$, and decoupling the resulting
quartic interaction using the $Q$ matrix.  For $\eta=0$, this action
has a non-compact $O(n,n)$ symmetry, reflecting the physical
equivalence of replicas.  In many respects, $\eta$ acts as a symmetry
breaking field, but is also essential to obtain convergence of the
functional integral.  Physical correlators, which are derived from the
logarithm of $Z_0$ ($\ln Z_0 = \lim_{n\rightarrow 0} (Z_0^n-1)/n$),
must be computed in the limit $n \rightarrow 0$.  Writing $Q =
\left(\matrix{Q_{++} & Q_{+-} \cr Q_{-+} & Q_{--}}\right)$, the
density of states is, e.g., $\rho = (1/\pi){\rm Im} \langle
Q_{++;\alpha\alpha} \rangle$, where no sum is implied over the
(arbitrary) replica index $1 \leq \alpha \leq n$.  Up to non-singular
contact terms, the correlator ${\cal D}({\bf r};0) = \langle
Q_{+-;\alpha\alpha}(\bbox{r}) Q_{-+;\alpha\alpha}(\bbox{0})\rangle$.

The ladder approximation is recovered as the leading term in a
saddle point expansion of the effective action for $Q$, obtained by
integrating out the bosonic fields.  At the mean-field level,
$\bar{Q}_{++} = - \bar{Q}_{--} = i/(2va)$,
reflecting the finite density of
states.

Expanding $Q = \bar{Q} + \tilde{Q}$, the $\tilde{Q}_{++}$ and
$\tilde{Q}_{--}$ fields have only massive fluctuations around the
saddle, but the $Q_{+-}$ and $Q_{-+}$ fields are massless Goldstone
bosons for $\eta=0$.  Following the conventional non-linear sigma model
approach, we ignore the massive fluctuations and focus on the
Goldstone modes, staying within the saddle point manifold by
eliminating $Q_{++}$ and $Q_{--}$ via the zero momentum, tree
equations of motion.  The resulting theory may be expressed solely in
terms of the off-diagonal submatrices of $\tilde{Q}$, and is governed
by an action of the form
\begin{eqnarray}
S_{\rm eff} & = & {1 \over {2\pi\rho}}\int_{\bbox{p}} (-ivp_x + D p_z^2) {\rm
Tr} [Q_{+-}(\bbox{p})Q_{-+}(-\bbox{p})] \nonumber \\
& + & \prod_{i=1}^4\int_{\bbox{p}_i} \Gamma(\bbox{p}_i) {\rm Tr}
[Q_{+-}Q_{-+}Q_{+-}Q_{-+}],
\label{Qaction}
\end{eqnarray}
to $O(Q_{\pm\mp}^4)$.  The interaction vertex $\Gamma(\bbox{p}_i) =
\delta(\sum\bbox{p}_i) f(\bbox{p}_i)$, with $f(\bbox{p}) \sim c_1 p_x
+ c_2 p_z^2$ for small momenta, and all higher vertices must have
similar momentum structure owing to the continuous $O(n,n)$ symmetry.
Truncating at quadratic order leads to the ladder result for ${\cal
D}$, while retaining the anharmonic corrections is analogous to the
usual expansion of the appropriate non-linear sigma model.

With this formulation, the stability of the ``diffusive'' metal can be
evaluated from simple power counting.  To make the quadratic action
dimensionless, we rescale $z \rightarrow b z$, $x \rightarrow b^2 x$,
and $Q \rightarrow b^{-1/2}Q$ (with $b>1$ to focus on long-wavelength
behavior).  Under this transformation the quartic coupling amplitude
$\Gamma_0 \rightarrow \Gamma_0/b$, implying the {\sl irrelevance} of
anharmonic terms in $Q_{+-}$ and the stability of the metallic phase.
This is in contrast to the usual case of a non-chiral 2d
dirty conductor, in which the metallic fixed point is marginally
unstable.
Physically, the absence of
localization effects may be attributed to the ballistic motion along
the $x$ axis, which suppresses multiple scattering.  Although it
is possible that strong enough
disorder could
induce some kind of transverse localization, we have been
unable to find evidence of such a phase, and suspect that metallic behavior
persists for all
impurity concentrations.

In other 2d
electron models which exhibit diffusion at $T=0$,
the de-localized wavefunctions typically exhibit multi-fractal
scaling\cite{Bodo}.
A classic example is the IQHE plateau transition, which manifests
multi-fractal behavior via
anomalous
scaling of ${\cal D} (p,\Omega)$, in the limit
$Dp^2 << \Omega$.
But in the present case ${\cal D} (p,\Omega)$
scales trivially with $\Omega$ and $p_z$, implying
that the wavefunctions are {\it not}
multi-fractal.

Having established the stability of the metallic phase in the
non-interacting case, we turn to the combined influence of interactions and
disorder.  As shown by Altschuler and Aronov
(see Ref.~\onlinecite{Lee}, and Refs.  therein),
interaction effects are much enhanced by diffusive
motion in conventional metals.  For example, diffusive
relaxation of density fluctuations
leads to singularities in the tunneling density of states.  For our
semi-ballistic metal, one expects this effect to be absent, since
any charge buildup is
swept away at the chiral Fermi velocity.
Indeed, one can show that the diagrams responsible for the singularity
yield only smooth contributions in the chiral case.

The z-axis transport for
the surface sheath in the IQHE behaves much as a conventional
metal.  Dramatically different results are obtained for fractional filling
factors, as we now describe.  We focus on the odd-denominator fractional
states, particularly $\nu = 1/3$.
The edge states are then
chiral Luttinger liquids of charge $\nu$ Laughlin
quasiparticles\cite{Wen}.
Because these quasiparticles have integrity only within a single quantum Hall
plane, inter-layer transport must involve the tunneling of physical
{\sl electrons}, or equivalently $1/\nu$ quasiparticles.
The surface ``sheath" can be described using the bosonized Euclidean
action
\begin{eqnarray}
S & = & \sum_i \int_{-\infty}^\infty \! dx \int_0^\beta d\tau \bigg\{
{1 \over {4\pi\nu}} \partial_x\phi_i (i\partial_\tau\phi_i + v
\partial_x\phi_i) \nonumber \\ & & - t\cos[(\phi_i
-\phi_{i+1})/\nu]\bigg\},
\label{onethirdaction}
\end{eqnarray}
where $\phi$ is a boson field, and $\beta = 1/T$.  The charge density
is $n_i(x) = \partial_x\phi_i/(2\pi)$, and the operator $e^{i\phi(x)}$
creates a quasiparticle ($2\pi$ soliton in $\phi$) at position $x$.
The cosine term describes inter-layer tunneling with amplitude $t$.
In the ideal $\bbox{H} \parallel \hat{\bbox{z}}$ geometry, no flux
penetrates between successive edge modes; oscillatory phase factors
are therefore absent from this term.  To study the effects of
tunneling, we employ standard renormalization group methods
perturbative in $t$.  After introducing a cutoff $\Lambda$ in $q_x$,
one integrates out modes with $\Lambda e^{-d\ell} < |q_x| < \Lambda$.
Upon rescaling $x \rightarrow e^{d\ell}x$ and imaginary time $\tau
\rightarrow e^{d\ell}\tau$, the quadratic action is brought back to
its original form.  The tunneling amplitude is, however,
renormalized\cite{MPAF_edge}\
\begin{equation}
{{dt} \over {d\ell}} = (2 - 1/\nu)t,
\end{equation}
and is technically an irrelevant perturbation.
To extract the z-axis conductivity
imagine renormalizing until the rescaled temperature is of order the quantum
Hall gap $E_g$, which implies a rescaling factor $e^\ell = E_g/T$.  From
scaling one
then obtains an inter-layer tunneling time
$\tau_{in} \sim \tau_{\phi} [E_g/t(\ell)]^2$,
with $t(\ell) \sim t (T/E_g)^{1/\nu -2}$.
Here $\tau_{\phi} \sim 1/T$
is a thermal de-phasing time .  Since $t(\ell)$ scales to zero
at low temperatures,
$\tau_{in} >> \tau_{\phi}$ and the inter-layer
transport is incoherent.  The z-axis conductivity
can then be obtained from the diffusion constant $D \sim a^2/\tau_{in}$
and the Einstein relation $\sigma_{zz} = e^2 \rho D$.
This gives
$\sigma_{zz} \sim (e^2/h)(a/v\tau_{in}) = (const)T^{2/\nu-3}$,
which {\it vanishes} as $T \rightarrow 0$.

We now summarize our predictions for transport and thermodynamic
quantities.  For temperatures well below the bulk QHE gap, the surface
sheath dominates the z-axis transport.  In this regime, for $\nu=1$, a
temperature independent (sheet) conductivity is predicted:
$\sigma_{zz}(\nu=1) \rightarrow (e^2t^2\tau)/2\pi v$.  Each additional
full Landau level will give another surface ``sheath", which
contributes additively to the sheet conductance, so
$\sigma_{zz}(\nu=N) \approx N\sigma_{zz}(1)$ (see Fig.~2).  This
behavior is roughly consistent with recent data on
(TMTSF)$_2$ClO$_4$\cite{Chaikin}.  As discussed above, the $\nu=1/3$
state, if present, has a vanishing z-axis conductivity,
$\sigma_{zz}\sim T^3$.  In contrast to the quantized Hall plateaus,
the ``plateaus" in $\sigma_{zz}$ are unquantized, and will probably
exhibit small non-vanishing slopes even at the lowest temperatures, as
depicted in the Figure.  However, the ``step-like'' features, which
arise from the 3d phase transitions between successive bulk Hall
phases, can be sharp at $T=0$.  The low temperature behavior of the
``steps" depends upon the order of the bulk phase transition.  In
(TMTSF)$_2$X, these transitions might be first order, with the SDW
period changing discontinuously.  In this case, we expect a
discontinuous jump in $\sigma_{zz}$.  However, experience with the
usual QHE leads us to expect {\sl continuous} critical behavior in 2d
electron gas multi-layers; sufficient disorder may also drive the SDW
transitions second order.  In this case, the ``steps" will exhibit
critical singularities, as we now describe.

\begin{figure}
\vspace{15pt}
\caption{Predicted behavior of $\sigma_{zz}$.  We have indicated a
small with $\delta B$, such as would be present at finite temperature
in the case of an isotropic bulk critical point.}
\label{fig2}
\end{figure}

A continuous bulk plateau transition is worthy of study in its own
right.  Here we restrict the discussion to scaling properties which
influence the surface sheath.  Near the transition one expects an
(in-plane) localization length which diverges as $\xi \sim |\delta
B|^{-\nu}$.  Allowing for anisotropic scaling, the z-axis localization
length is assumed to vary as $\xi_z \sim \xi^\zeta$ (we expect $\zeta
\leq 1$ on physical grounds), whereas characteristic energy scales
vanish as $\omega \sim \xi^{-z}$ with $z$ a dynamical exponent.
Following standard methods,\cite{scalingref}\ we find that the bulk
conductivity $\sigma_{zz}^{\rm Bulk} \sim \xi^{\zeta-2} \Xi(T\xi^z)$.
This form connects activated behavior away from the transition (with
energy scale $T_0 \sim E_g (\ell_B/\xi)^z$, where $\ell_B$ is the
magnetic length) to the power-law temperature dependence
$\sigma_{zz}^{\rm Bulk} \sim T^{(2-\zeta)/z}$ at the critical point.
As an aside, we point out that $\sigma_{xx} \sim T^{\zeta/z}
\rightarrow 0$ at the 3d critical point, in contrast with the
prediction of a universal finite value in 2d.

Upon approaching such a continuous bulk transition, the surface sheath
develops a thickness $\xi \gg \ell_B$, and we
may estimate the z-axis sheet conductivity as $\sigma_{zz} \sim
\sigma_{zz}^{\rm Bulk} \xi$.  Using the bulk scaling results, this
implies singular behavior at zero temperature for the ``steps" in
$\sigma_{zz}$:
\begin{equation}
\sigma_{zz}
\sim \xi^{\zeta-1} \sim |\delta B|^{(1-\zeta)\nu} .
\end{equation}
Experience with other similar disordered systems suggests that the
critical behavior may be isotropic, so that $\zeta=1$\cite{YoungVG}.
If so, we again find a jump in $\sigma_{zz}$.  A jump due to an
isotropic but continuous bulk transition can be differentiated from a
bulk first order transition by its behavior at finite temperature: in
the former case, the step develops a width $\delta B \sim T^{1/(z \nu
)}$, while for a first order transition the discontinuity should be
present even at $T \ne 0$.

The SDW compounds provide a unique opportunity to study the specific
heat of quantum Hall edge modes, in this case those of the whole
sheath.  In 2d electron gases, the edge contribution to $C$ is masked
by that of the localized states in the interior.  While this should
also be the case in multi-layers, the SDW gap implies that the bulk
contribution to $C$ is activated at low temperatures.  One can then
measure the specific heat from the gapless surface modes.  They
contribute a linear temperature dependence to the specific heat per
area: $C_{A} = N (\pi k_B^2 T/6\hbar a v)$.  Since the ratio $C/T$ is
linear in the number $N$ of full Landau levels, one expects ``steps"
in this quantity as well.  Recent specific heat data in the $ClO_4$
Bechgaard salt appears roughly consistent with this variation.  The
evidence for a latent heat peak near the transitions is less
clear\cite{Scheven}.

Finally, we consider the effects of a transverse magnetic field ($B_y
\neq 0$) on the integer surface sheath, induced (albeit non-uniformly)
by tilting the sample.  In the gauge $A_x = -B_y z$, this field shifts
the relative Fermi momenta between edge modes in adjacent layers by
$\delta k_F = eaB_y$.  In the absence of impurities, momentum
conserving inter-layer tunneling is not possible below an energy
scale, $\Delta_B \sim vea B_y$.  This leads to activated z-axis
transport, $\sigma_{zz} \sim \sigma_{zz,0} \exp(-\Delta_B/T)$,
characteristic of decoupled 1d conductors.  With impurity scattering,
a finite conductivity will be restored at $T=0$.  But for a relatively
clean surface, a {\it positive} magneto-{\sl resistance} for z-axis transport
is expected.

We are grateful to P.  Chaikin for helpful conversations.  This work has
been supported by the National Science Foundation under grants No.
PHY94--07194, No.  DMR--9400142 and under the Alan T. Waterman
award grant No. DMR-9528578.

\end{document}